\documentclass[useAMS,usenatbib]{mn2e}
\usepackage{graphicx}
\usepackage{amssymb}

\def\mic {\hbox{$\mu$m}}

\def\eq#1{\begin{equation} #1 \end{equation}}
\def\Ch {{\'{c}}}

\title[Inner regions of protoplanetary discs ]{Inner dusty regions of protoplanetary discs -- I. High resolution temperature structure}
\author[D. Vinkovi\Ch]{Dejan Vinkovi\Ch$^{1}$\thanks{E-mail: vinkovic@pmfst.hr}\\
$^{1}$Physics Department, University of Split, Nikole Tesle 12, HR-21000 Split, Croatia}

\begin{document}

\date{Draft - October 2011}

\pagerange{\pageref{firstpage}--\pageref{lastpage}} \pubyear{2011}

\maketitle

\begin{abstract}

Our current understanding of the physical conditions in the inner regions of protoplanetary discs is becoming increasingly challenged by the more detailed observational and theoretical explorations. Calculation of dust temperature is one of the key features we strive to understand and a necessary step in image and flux reconstruction. We explore coexistence of small ($0.1\mu m$ radius) and big ($2\mu m$ radius) dust grains can coexist at distances from the star where small grains would not survive without big grains shielding them from the direct starlight. The study required a high resolution radiative transfer calculation capable of resolving large temperature gradients and disc surface curvatures caused by dust sublimation. The calculation was also capable of resolving temperature inversion effect in big grains, where the maximum dust temperature is at visual optical depth of $\tau_V\sim 1.5$. We also show disc images and spectra, with disentangled contributions from small and big grains. Big grains dominate the near IR flux, mainly because of the bright hot inner disc rim. Small grains populate almost the entire inner disc interior, but appear in the disc surface at distances $2.2$ times larger than the closest distance of big grains from the star. Nevertheless, small grains can contribute to the image surface brightness at smaller radii because they are visible below the optically thin surface defined by stellar heating. Our calculations demonstrate that the sublimation temperature does not provide a unique boundary condition for radiative transfer models of optically thick discs. The source of this problem is the temperature inversion effect, which allows survival of optically thin configurations of big grains closer to the star than the inner radius of optically thick disc. Future attempts to derive more realistic multigrain inner disc models will require the numerical resolution shown in our study, especially if dust dynamics is considered were grains can travel through zones of local temperature maxima.

\end{abstract}


\section{Introduction}

A pressing issue raised by the ongoing proliferation of extrasolar planet discoveries is identifying conditions under which planets form and develop. A general scenario invokes a circumstellar (protoplanetary) disc of dust and gas around young pre-main-sequence stars, where planets are expected to form. The advancements in observational and theoretical techniques in recent decades enabled extensive exploration of the evolution of these discs \citep{Williams}. One of the key issues in protoplanetary disc exploration is understanding the evolution of dust properties. A major observational contributor to this field are investigations of dust properties deduced form dust infrared (IR) excess in the spectral energy distribution (SED) \citep[e.g. various recent results from {\it Spitzer} surveys, such as][etc. and references therein]{Sargent09, Watson09, Juhasz10, McClure10, Oliveira10, Manoj11, Oliveira11}. However, the exact details of the spatial disc structure are difficult to determine from the SEDs alone due to the intrinsic mathematical degeneracy of models reproducing only flux measurements \citep{Vinkovic03}. This makes direct imaging of protoplanetary discs an indispensable tool in exploration of their properties, but it also poses difficult challenges for theoretical models \citep[e.g.][]{Millan-Gabet07, DiskImaging, ReviewARAA, Williams}.

The innermost disc regions, within a few AU around the central star, have attracted a special attention in this field of research, since we expect formation of terrestrial type of planets in this disc zone. It is also a zone of dust sublimation, high gas densities and complex dust dynamics \citep[e.g.][]{Ciesla09,Vinkovic09,Hughes10,Turner,Armitage}. At the same time, it is a very difficult observational target due to its small scale and proximity to the central star \citep{Millan-Gabet07}. A strong impetus to the theoretical exploration of inner disc regions was given by development of semi-analytical models inspired by the observed properties of the near IR part of SEDs \citep{ReviewARAA}. This is a part of the spectrum where the inner disc emits thanks to its most resilient dust grains that survive up to temperatures of 1500-2000K.

In recent years there is a tendency to improve upon the existing models. One motivation for that comes from observations showing various features that cannot be accommodated by the existing simplified and time-steady models \citep{ReviewARAA}. Another motivation for improvements spurs from theoretical investigations into the inner disc radiative transfer, grain and gas composition and disc geometry. For example, truncation of the inner disc by dust sublimation results in a curvature of inner disc rim \citep{Isella05,Tannirkulam07}; multigrain dust models result in a complicated geometries and density distributions \citep{Kama}; big grains are gray in the near IR and experience maximum temperature within the dust cloud and not at the very surface exposed to the starlight \citep{Vinkovic06,Kama}; dust dynamics is influenced by radiative pressure and magnetic fields in complicated ways \citep{Vinkovic09,Turner}; dust and gas are thermally decoupled in optically thin regions, which makes gas hotter than initially expected \citep{Thi}; gas itself can be a significant  source of near IR flux and/or optically thin dust survives closer to the star than the optically thick inner disc rim \citep{Tannirkulam08,Benisty10}.

In this series of papers, we will investigate various theoretical and numerical improvements in modelling  of the inner disc structure. We start with a study of improved numerical reconstruction of dust thermal structure under a condition of multigrain dust composition. We use passively heated discs as a necessary initial step in identifying the most basic radiative transfer effects. Our focus is on coexistence of small and big dust grains, because these two dust properties yield big differences in dust sublimation behaviour. When used alone, small grains sublimate at considerably larger distances from the star than big grains. However, when mixed together they can coexist at distances dictated by big grains, but quantitative details are still not known. Another problem arises from a complicated behaviour of big grains' temperature under direct exposure to stellar heating. As mentioned above, temperature initially increases with optical depth, reaches its maximum at some optical depth, and then decreases as we move deeper into the optically thick disc interior. Since dust sublimation is typically used as a unique boundary condition for inner disc structure, we will explore how solid is that assumption under high resolution numerical reconstruction of multigrain dust temperature.
In section \ref{sec_radtrans} we describe some caveats in numerical treatment of the problem; in section \ref{sec_results} we described obtained temperature structure and disc images and spectra; and in section \ref{sec_discuss_conclude} we discuss implications of our findings.

\section{Radiative transfer modelling }
\label{sec_radtrans}

\subsection{Description of the model}

\begin{figure}
  \includegraphics[width=3.4in]{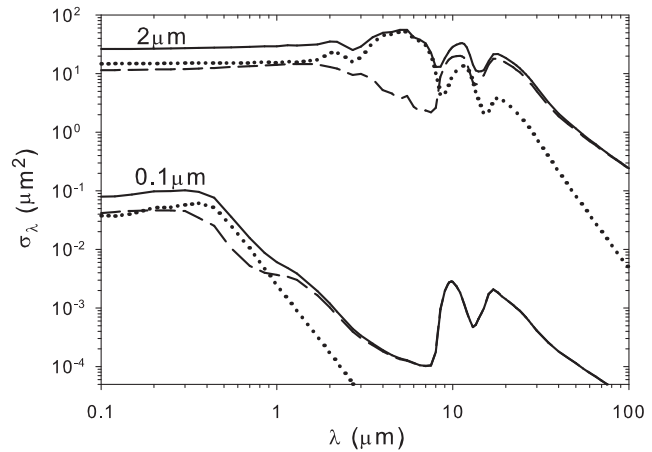}\\
  \caption{Cross section of olivine dust grains \citep{Dorschner} of $0.1\mu m$ and $2\mu m$ radius used in modelling. The dashed lines represent emission, the dotted lines scattering and the solid lines total extinction cross section.}\label{Fig_cross_sec}
\end{figure}

The main focus of this work is on the high resolution radiative transfer and analysis of dusty disc surface.
Hence, we simplify the problem of vertical disc structure and use the model developed by \cite{ShakuraSunyaev}, where the gas and dust are well mixed and the number density of dust particles of type $\alpha$ in the cylindrical coordinate system $(\varrho,z)$ is
 \eq{\label{n_alpha}
  n_\alpha(\varrho,z) = N_\alpha\varrho^{-p}\exp\left({-h_0\frac{z^2}{\varrho^{2m}}}\right),
 }
 with parameters $p=2$, $m=1.25$ and $h_0=1800$. The coordinates $(\varrho,z)$ used in our calculations are scaled with the smallest distance of dust from the star, which we call the inner disc radius $R_{in}$ and it is located in the disc midplane. This distance is not known in advance and it is a part of the output from radiative transfer calculation. The outer disc radius is fixed to $R_{out}=100R_{in}$. The number density of dust grains $\alpha$ at $\varrho=1$ and $z=0$, which is the location of radius $R_{in}$, is equal to $N_\alpha = n_\alpha(1,0)$. We use olivine dust opacities from \cite{Dorschner} and two coexisting grain types $\alpha$ of $0.1\mu m$ and $2\mu m$ radius. Their cross section is shown in Fig.~\ref{Fig_cross_sec}. The ratio of their number densities at $(1,0)$ is fixed to $N_{0.1}:N_{2}=10^4:1$. The dust sublimation temperature is $T_{sub}=1500$K. The spectral shape of the stellar radiation is taken from the \cite{Kurucz} model atmosphere for a $T_*=10\,000$K star.

Instead of fixing the absolute value of dust densities, our radiative transfer calculation works with a fixed optical depth along an arbitrary radial line. We use visual optical depth $\tau_{_V}=10^4$ at $0.55\mu m$ toward the star in the disc midplane. This optical depth is a line integral of dust visual cross sections ($\sigma_{_{0.1,V}}$ and $\sigma_{_{2,V}}$) and dust number densities ($n_{_{0.1}}$ and $n_{_2}$), but we do not know in advance the integral limits because our two dust types sublimate at different distances from the star. We expect $2\mu m$ grains to survive closer to the star than $0.1\mu m$ grains because they are better emitters in the near IR (see Fig.~\ref{Fig_cross_sec}), where $T_{sub}$ has its spectral maximum. This means that $2\mu m$ grains are cooled more efficiently and exist at $R_{in}$, while the sublimation front for $0.1\mu m$ is not known in advance.

We will see in the end that in our optically thick example smaller grains survive very close to $R_{in}$ thanks to bigger grains shielding them from the direct stellar radiation. This simplifies the optical depth integral
 \eq{
 \tau_{_V} = \int\limits_{\approx 1}^{R_{out}/R_{in}} (\sigma_{_{0.1,V}}n_{_{0.1}}(\varrho,0)+\sigma_{_{2,V}}n_{_2}(\varrho,0))R_{in}d\varrho
 }
and we can derive the dust density scale using equation \ref{n_alpha}
 \eq{\label{N01}
 N_{0.1} = { \tau_{_V} \over \left(\sigma_{_{0.1,V}} +  \sigma_{_{2,V}} N_{0.1} / N_{2} \right) \left( 1 - R_{in}/R_{out}\right) R_{in}} .
 }
This equation shows that density scale is directly related to the distance scale $R_{in}$. Using values for our parameters in equation \ref{N01} gives $N_{0.1} = 3.76\times 10^8 / (R_{in}/R_\odot) m^{-3}$. Since we do not know $R_{in}$ in advance, we also do not know the absolute density scale.

To solve the radiative transfer problem we only need to specify $T_{sub}$ and $T_*$, the optical depth $\tau_{_V}$, the spectral shape of dust cross sections and the shape of stellar spectrum. All other properties can be expressed in dimensionless terms. This is a consequence of intrinsic degeneracy of radiative transfer equations, as shown by \cite{IE97} in their derivation of general scaling properties of the radiative transfer problem for radiatively heated dust. Luminosity $L_*$ is irrelevant and it enters as a parameter only after we obtain $R_{in}$ in units of stellar radii $R_*$ from the output of radiative transfer calculation. Then we can scale our model according to the stellar radius derived from the luminosity: $R_*=R_\odot(T_\odot/T_*)^2(L_*/L_\odot)^{0.5}$.

\begin{figure}
  \includegraphics[width=3.4in, angle=180]{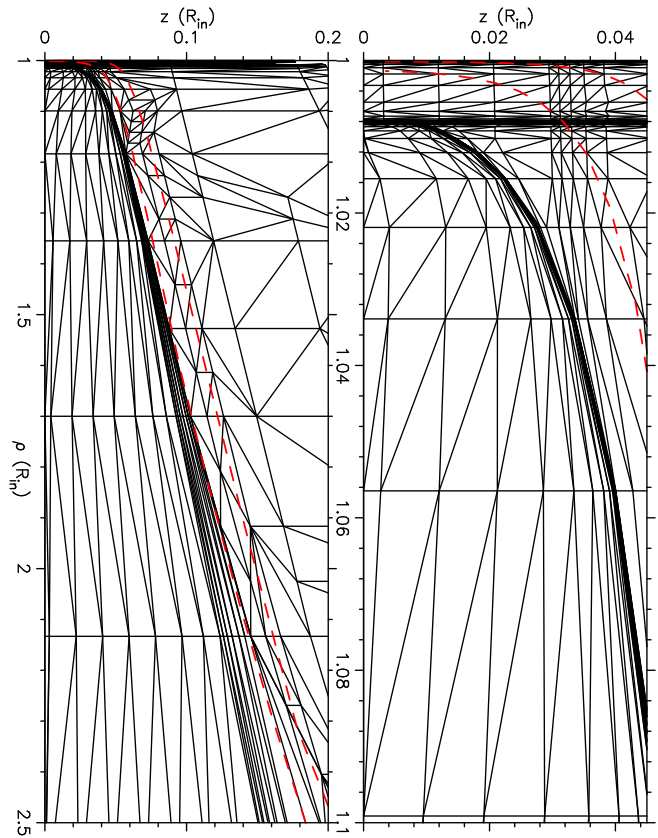}\\
  \caption{Computational grid used in the final model. The panels show two spatial scales of the inner disc region in cylindrical coordinates $(\varrho,z)$. Thick dashed lines are radial visual optical depths of $\tau_V=0.1$ and $\tau_V=1$. A section of the grid with increased resolution traces the surface of disc populated with $0.1\mic$ grains, while $2\mic$ grains exist at all grid vertices.}
  \label{Fig_grid}
\end{figure}

\subsection{Boundary condition at the inner rim}

The parameter that has been traditionally used as a boundary condition for the inner rim of dusty discs is $T_{sub}$. The physical motivation for this was the expectation that the dust temperature of externally heated dusty clouds always monotonically decreases with the increasing distance from the heat source. This means that the maximum dust temperature is achieved at the closest distance to the star that a dust particle can reach before sublimating away.

However, this concept was found incorrect in the case of externally heated optically thick clouds made of big ($\gtrsim 1\mu m$) dust grains. Such grains have an ability to efficiently absorb the diffuse IR radiation originating from the cloud's interior, which results in an increase of temperature within the dust cloud at the visual optical depth of $\tau_{_V}\sim 1$, instead of on the very surface exposed to the stellar heating. This process of temperature inversion is similar to the ''greenhouse effect'' and it was first discovered numerically in single grain size models \citep{Dullemond02,Isella05,VIJE} and then proved analytically for multigrain dust mixtures \citep{Vinkovic06}. The analytical analysis showed that big grains dominate the cloud surface, while smaller grains can exist immediately behind the zone of temperature inversion. The analysis also showed that the effect exists for both types of radiative transfer boundary conditions: (1) a constant external flux heating the cloud, with no limits on the dust temperature, and (2) a fixed maximum dust temperature, corresponding to dust sublimation.

The inner regions of protoplanetary discs provide ideal conditions for emergence of the temperature inversion phenomenon. Accretion, which is ubiquitous to protoplanetary discs, constantly resupplies the inner regions with big grains that grow in the disc. Hence, the optically thick discs should be in a permanent process of minimization of their inner disc radius, with big grains populating the inner disc surface and dictating the radiative transfer. A complication arises when we try to deal with the spatial scale of the disc surface because the surface is defined by the optical depth of $\tau_{_V}\lesssim 1$ as a zone where the most of stellar radiation is absorbed. This optically thin zone is, therefore, not necessarily geometrically thin. Instead, it could extend much closer to the star, creating a large optically thin dusty zone spreading over radii smaller then the inner radius of optically thick disc. Since dust can exist only up to its sublimation temperature, \cite{Vinkovic06} proposed using two inner radii that discriminate between these two coexisting, optically thin and thick, disc zones:
 \eq{\label{Rin_thin_thick}
   R_{in}=\frac{\Psi R_*}{2}\left( \frac{T_*}{T_{sub}} \right)^2,
 }
where $\Psi=2$ is used for the optically thick \citep{DDN} and $\Psi\sim 1.2$ for the optically thin inner radius \citep{VIJE}. Note that here we assume that the disc opacity is dominated by big grains that have gray opacity in the near IR. Smaller grains are inefficient emitters of the near IR radiation and, therefore, overheat and sublimate away at these distances form the star. Smaller grains can survive either at larger distances from the star or hidden in the disc interior behind gray dust.

\begin{figure*}
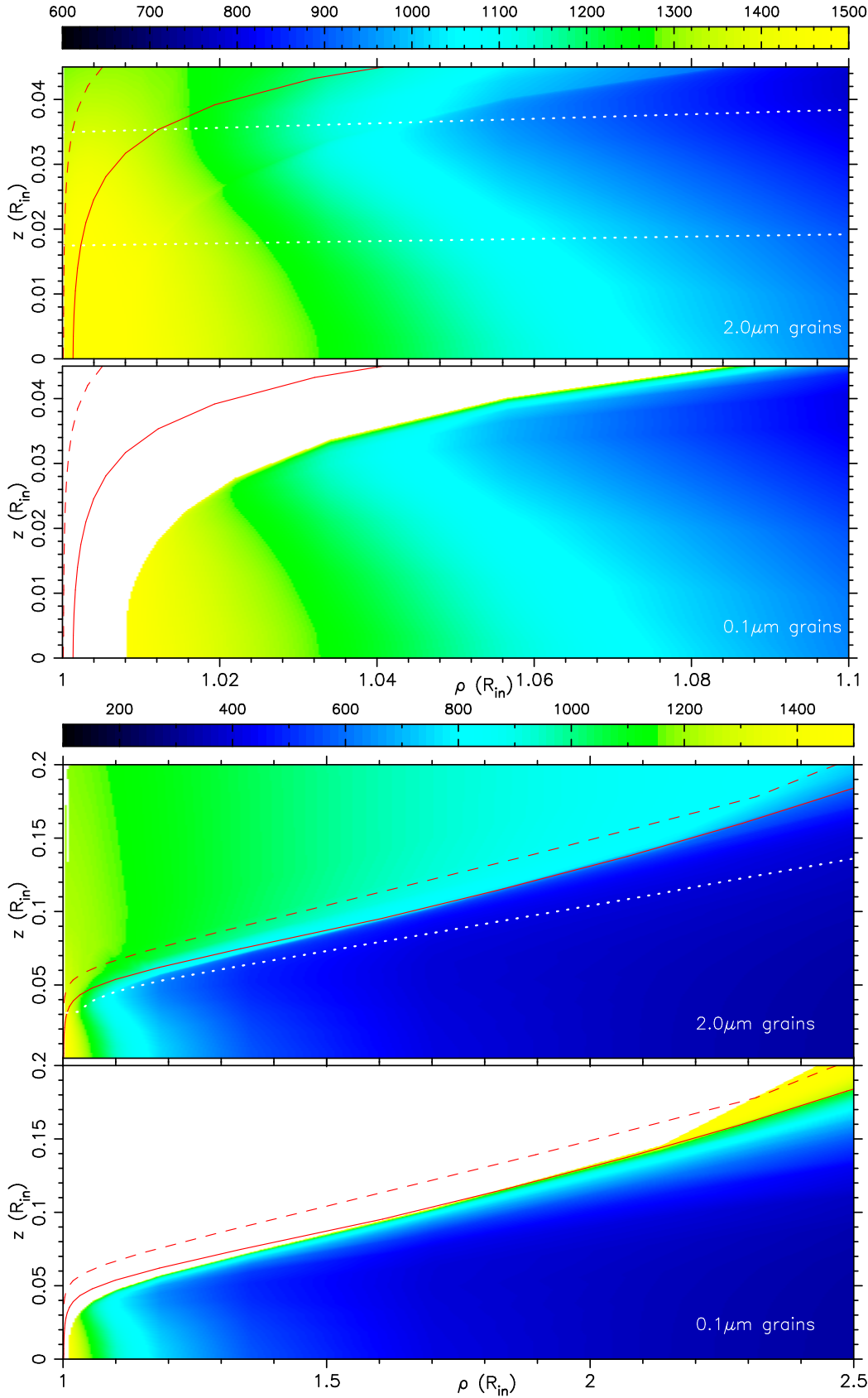

  \includegraphics[width=4.4in, angle=-90]{Figure_03a.ps}\\
  \includegraphics[width=4.4in, angle=-90]{Figure_03b.ps}\\
  \caption{Temperature of dust grains shown at two spatial scales in cylindrical coordinates for two coexisting grain radii (indicated in images). The dashed line indicates the radial optical depth of $\tau_V=0.1$, while the solid line is $\tau_V=1$. Note how $0.1\mic$ grains survive in the optically thin disc surface only at radial distances of $\varrho \gtrsim 2.2$, but in the disc interior they exist very close to $R_{in}$. The dotted lines in the most upper panel indicate radial lines of $88^\circ$ and $89^\circ$ inclination angle used in Fig.~\ref{Fig_T_tauV} and \ref{Fig_T_R}. The dotted line in the third from the top panel shows the vertical optical depth of $\tau_V=1$.}
  \label{Fig_temperature2}
\end{figure*}

\cite{Kama} showed these effects numerically, using discs with various dust compositions and surface densities. The optically thick radius in their models dominated by big grains are in agreement\footnote{Small deviations from equation \ref{Rin_thin_thick} are visible in their models because they use dependence of dust sublimation on gas density, which slightly increases dust sublimation temperature at the inner disc rim.} with equation \ref{Rin_thin_thick}. The optically thin zones in their models are dictated by predefined power law density profiles and, therefore, do not extend to the theoretical minimum distance to the star.

Models by \cite{Kama} demonstrate the boundary condition problem that we face when the spatial dust density distribution of the optically thin zone is free to vary. In order to simplify the problem and make it manageable, in this paper we do not explore properties of the optically thin zone. Our goal is limited to the exploration of temperature inversion structure in 2D and survival of small grains in the disc. We achieve this by fixing the maximum dust temperature {\it anywhere} in the disc to $T_{sub}$. This means that we do not know the dust temperature $T_{in}$ at $(\varrho=1,z=0)$, but we expect it to be smaller than $T_{sub}$.

Such a boundary condition is easy to postulate, but it creates substantial numerical challenges, especially in a case of high resolution temperature calculation like ours. Our numerical approach was semi-autonomous, where the computational grid updates and $T_{in}$ corrections where checked by hand for precision and error control. We know in advance that $2\mic$ grains in our model should survive at all grid points. On the other hand, $0.1\mic$ grains must be removed from the surface of inner disc where these grains overheat. They survive only in the interior shielded by $2\mic$ grains, but their exact location of survival is not known in advance.

The solution is looked for by iterating between radiative transfer calculations and dust removal. These iterations were combined with corrections by hand of $T_{in}$. In addition to that, we updated the computational grid structure after each dust removal because our grid is irregular and automatically traces spatial and optical depth gradients. Even though the majority of this iterative process is automatic, we had to check each step by hand to avoid errors and grid imperfections. Nonetheless, small deviations from $T_{sub}$ are very difficult to suppress, but we managed to keep the error within 3\% (45K for $T_{sub}=1500K$).

\subsection{Radiative transfer calculation}

Numerical radiative transfer calculations were performed with the code LELUYA (www.leluya.org). It solves the integral equation of the formal solution of radiative transfer with axially symmetric dust configurations, including dust scattering, absorption, and thermal emission. The solution is based on a long-characteristics approach to the direct method of solving the matrix version of the integral equation \citep{Kurucz69}. The equations are solved on a highly unstructured triangular self-adaptive grid that traces simultaneously both the density gradients and the optical depth gradients over many orders of magnitude in spatial and optical depth space. The code is parallelized and written in C. Theoretical and computational details are explained by \cite{VinkovicPhD}.

The key for a high resolution temperature structure is the grid used in our calculations. The code creates a grid by integrating optical depth along lines that cut through the largest optical depth gradients. In our case these are radial lines and vertical lines parallel to the symmetry axis. The grid vertices are chosen on these lines according to the following equation:
 \eq{
  \frac{\triangle\tau}{\tau}+\frac{\triangle L}{L} = 2,
 }
where $\triangle\tau$ is the optical depth step, $\triangle L$ is the spatial step, where both are calculated relative to the last created grid vertex, and $\tau$ and $L$ are the optical depth and the spatial distance from the starting point of integration. If used carefully, with starting points in low density areas and at dust sublimation surfaces, this equation enforces higher grid resolution in zones of optical depth gradients. Vertices in optically deep interior are distributed more sparsely because the radiation field is slowly varying. In addition to this automatic grid generation, we also inspected the grid by eye and added extra grid vertices where we considered necessary. The final grid is shown in Fig.~\ref{Fig_grid}.

Each grid vertex contains an angular grid defining directions of integration of radiative transfer equation. It would be simple and fast to use always the same type of angular grid, but vertices on the disc surface are surrounded by an anisotropic radiation field that changes its geometry from vertex to vertex. Therefore, we use a method where a unit sphere is split self-adaptively into small spherical triangles according to the number of vertices visible through a triangle in 3D space. This enables increased resolution in directions of radiation field gradients. The surfaces of spherical triangles are used as corresponding statistical weights in the discretized angular integration.

\begin{figure}
  \includegraphics[width=3.4in]{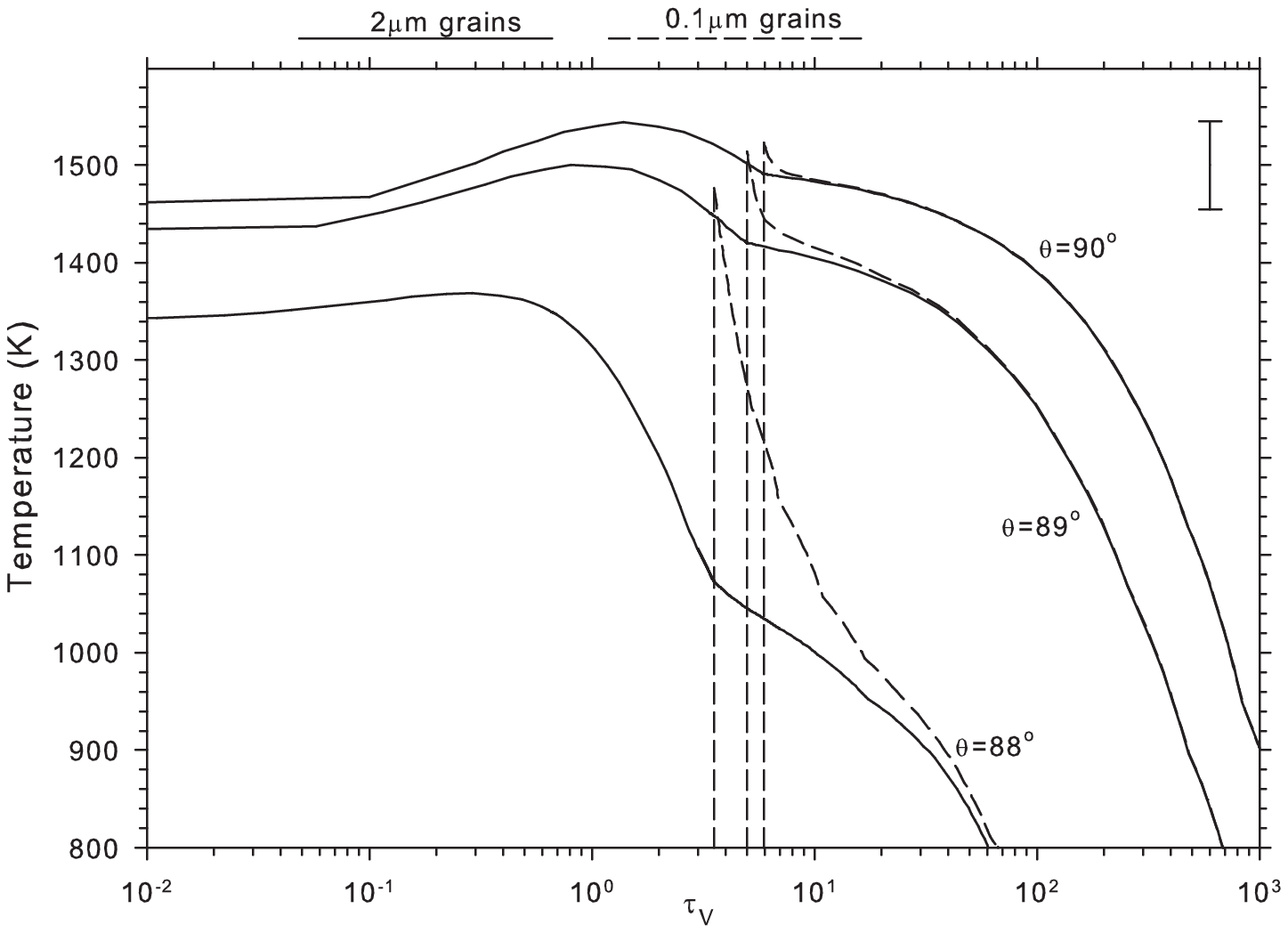}\\
  \caption{Temperature profiles along radial visual optical depth lines for different inclination angles $\theta$ relative to the symmetry axis (z-axis): $90^\circ$ (midplane), $89^\circ$ and $88^\circ$ (shown in Fig.~\ref{Fig_temperature2} as dotted lines in the top panel). The solid line indicates the temperature of $2\mic$ grains and the dashed line of $0.1\mic$ grains. The error bar marks 3\% error tolerance on the dust sublimation temperature of 1500K. }
  \label{Fig_T_tauV}
\end{figure}

\begin{figure}
  \includegraphics[width=3.4in]{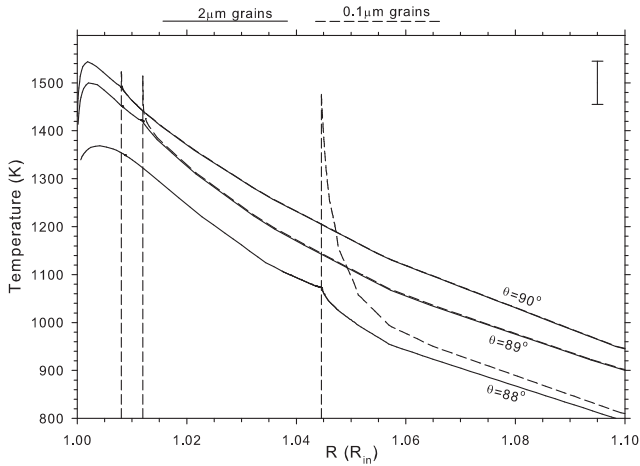}\\
  \caption{Same as Fig.~\ref{Fig_T_tauV}, except that the temperature is shown relative to the spatial scale $R=(\varrho^2+z^2)^{0.5}$ along radial lines. Steep decline of $0.1\mic$ grain temperature explains increased grid resolution in the sublimation zone of $0.1\mic$ grains (see Fig.~\ref{Fig_grid}). }
  \label{Fig_T_R}
\end{figure}

\begin{figure}
  \includegraphics[width=3.in]{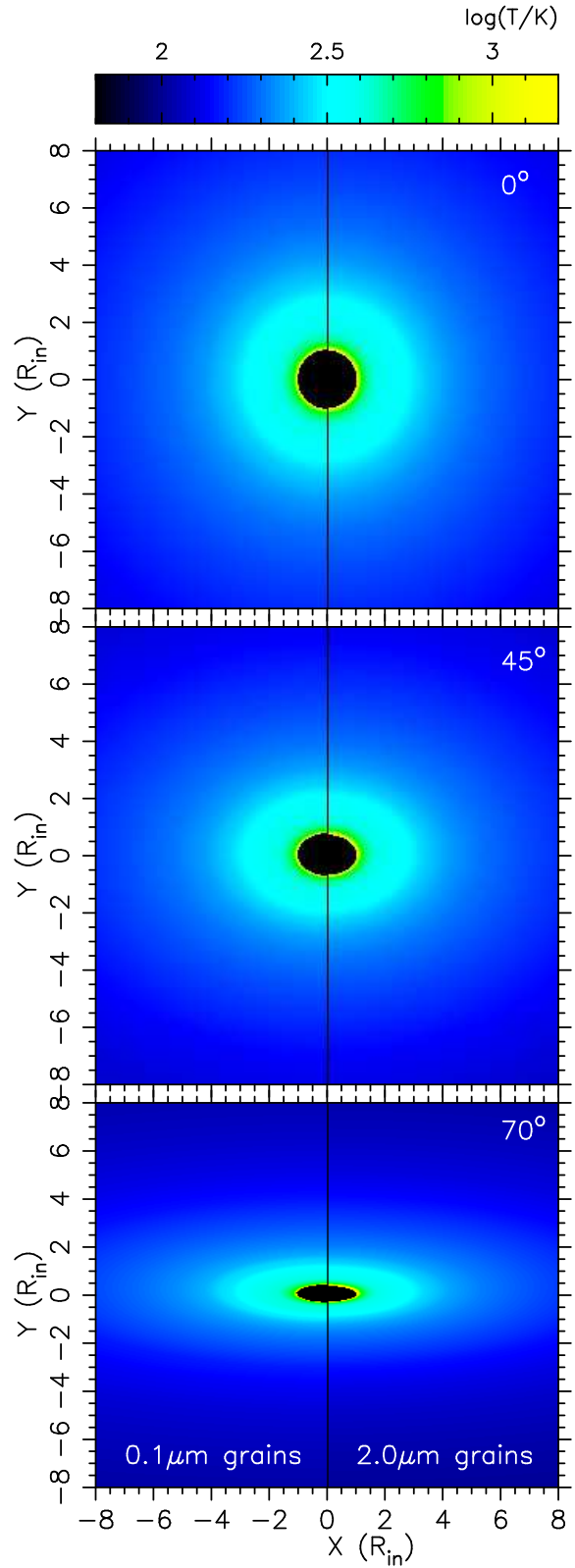}\\
  \caption{Map of temperature along the line of sight for different inclination angles. The images show temperatures at optical depth $\tau=1$ for $2.2\mic$ wavelength. The left side of images shows temperatures of $0.1\mic$ dust grains and the right side shows temperatures of $2\mic$ dust grains.}
  \label{Fig_image_2.2mic_temperature}
\end{figure}

\section{Results}
\label{sec_results}

We obtained the inner disc radius of $R_{in}=44.72R_*$, which agrees with equation \ref{Rin_thin_thick} for the case of optically thick dusty disc edge. In real units it corresponds to $R_{in}=0.0687(L_*/L_\odot)^{0.5}$AU for a 10,000K star. Typical luminosities of Herbig Ae stars are $L_*=20-100L_\odot$, hence, typical values for the inner optically thick disc radius are $R_{in}=0.31-0.69$AU.

\subsection{Temperature structure}

The dust temperature distribution is shown in Fig.~\ref{Fig_temperature2} for both $2\mic$ and $0.1\mic$ grains. The optically thin disc surface is populated only by big grains at distances $\varrho \lesssim 2.2$ from the star. However, smaller grains survive in the optically thick disc interior, where they are heated by diffuse infrared radiation and protected from the direct stellar radiation. This enables smaller grains to survive almost as close as $R_{in}$ from the star.

\begin{figure}
  \includegraphics[width=3.4in]{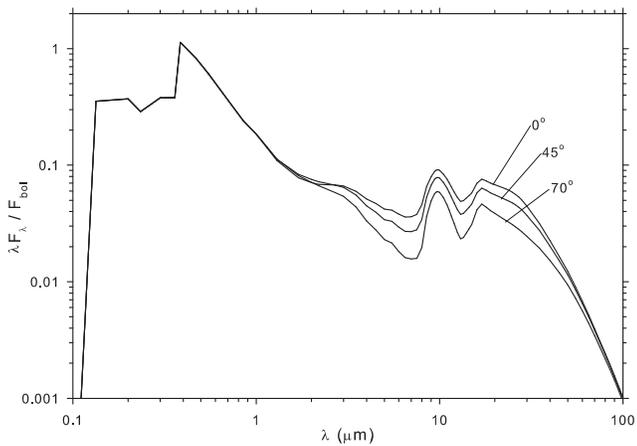}\\
  \caption{Spectral energy distribution of the model for three different disc inclination angles: $0^\circ$ (face on), $45^\circ$ and $70^\circ$.}
  \label{Fig_spectra}
\end{figure}

\begin{figure}
  \includegraphics[width=3.4in]{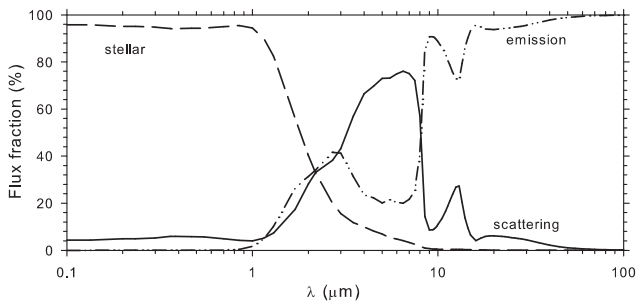}\\
  \caption{Wavelength variation of the relative contribution of stellar, emission and scattering flux to the total flux for $45^\circ$ disc inclination. Other inclinations have very similar relative contributions. Note the dominance of scattering in the near IR due to big $2\mic$ grains (see also dust cross sections in Fig.~\ref{Fig_cross_sec}).}
  \label{Fig_fractions}
\end{figure}

\begin{figure}
  \includegraphics[width=3.4in]{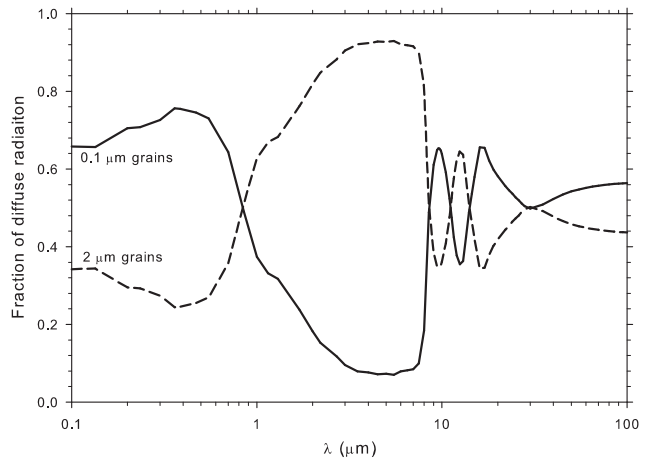}\\
  \caption{Wavelength variation of the relative contribution of $0.1\mic$ and $2\mic$ grains to the total observed diffuse flux for $45^\circ$ disc inclination. Other inclinations have very similar relative contributions. Note how $2\mic$ grains dominated the near IR because the opacity of $0.1\mic$ grains drops fast in this wavelength region (see Fig.~\ref{Fig_cross_sec}).}
  \label{Fig_fractions_grainsize}
\end{figure}

The temperature inversion effect is reproduced by our model, with the temperature maximum at $\tau_V\sim 1.5$ in the optical depth space and at $\sim 1.005R_{in}$ from the star. As a result of this effect, the temperature decreases very slowly with the optical depth and the zone of high dust temperature is not confined to a geometrically thin disc edge. With high temperature we mean temperatures high enough to quickly anneal the silicates \citep{Hill}, which is $\gtrsim$950K. In Fig.~\ref{Fig_temperature2} we can see that the disc temperature is above 950K between $R_{in}$ and $\sim 1.1R_{in}$, which is geometrically a significant part of the inner disc.

A more detailed insight into the temperature structure is given in Fig.~\ref{Fig_T_tauV} and Fig.~\ref{Fig_T_R}. They show cuts through the disc along radial lines of different angles $\theta$ relative to the symmetry axis (z-axis). Fig.~\ref{Fig_T_tauV} displays the dust temperature relative to the radial optical depth. In the midplane ($\theta=90^\circ$) temperature stays above the annealing limit of 950K almost until $\tau_V\sim 10^3$. Small grains sublimate at $\tau_V\sim 4-6$ and their temperature drops faster then the temperature of bigger grains, such that both grain sizes reach the same temperature at $\tau\gtrsim 10$. This is in agreement with the qualitative theoretical predictions by \cite{Vinkovic06}. The cut along $\theta=88^\circ$ shows a slower decline of $0.1\mic$ grain temperature, but this is because this cut goes through the disc region where the largest temperature gradient is perpendicular to the disc plane (see the upper dotted line in the top panel in Fig.~\ref{Fig_temperature2}).

Fig.~\ref{Fig_T_R} displays the same radial cuts, but relative to the spatial distance from the star. Here we can see that the dust temperature in the midplane stays above 950K almost up to $1.1R_{in}$. Small grains sublimate at distances from $1.008R_{in}$ (midplane) to $1.045R_{in}$ (for $\theta=88^\circ$). The temperature gradient for $0.1\mic$ grains is very steep between the point of $T_{sub}$ and the point of thermalization with $2\mic$ grains. This explains why the computational grid has such a high resolution in the region of $0.1\mic$ dust sublimation (see Fig.~\ref{Fig_grid}). The spatial step in this region goes as low as $\triangle L = 3.7\times 10^{-6}R_{in}$.

The temperature of $2\mic$ grains is slightly increased at the location of sublimation of $0.1\mic$ grains. The effect is small because the near IR emissivity of $0.1\mic$ grains is too small to provide a significant contribution to the local diffuse radiation. This would not be the case if the number density of bigger grains were reduced to the levels where the diffuse radiation is dominated by smaller grains. However, this would also mean that bigger grains do not dominate the optical depth and, in turn, cannot provide shielding to smaller grains. This would result in a larger inner disc radius dictated by the opacity of smaller grains.

Another way of observing the disc temperature structure is to focus on how deep we can see into the disc. For example, the dotted line in the third from the top panel in Fig.~\ref{Fig_temperature2} shows the optical depth of $\tau_V=1$ when the disc is viewed face on. This is approximately the depth of observable disc at $0.55\mic$ wavelength. Fig.~\ref{Fig_image_2.2mic_temperature} shows the dust temperature distribution at the optical depth of $\tau=1$ at $2.2\mic$ wavelength for different inclination angles. It reveals that we can see deep enough into the disc to sample both dust types even in the regions where smaller grains do not populate the disc surface.

\begin{figure*}
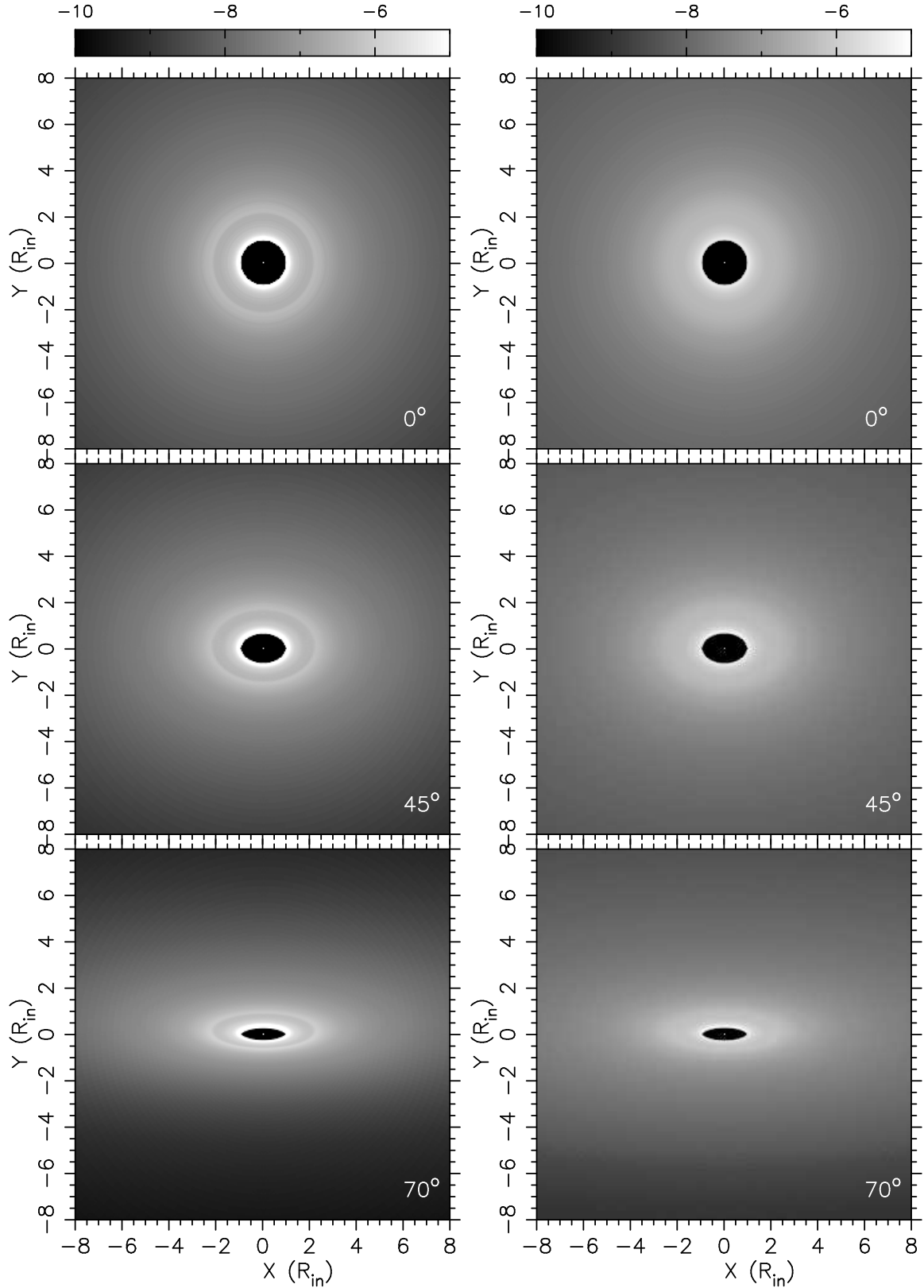

  \includegraphics[width=3in]{Figure_10a.ps}
  \includegraphics[width=3in]{Figure_10b.ps}\\
  \caption{Model images at $2.2\mic$ (left column) and $10\mic$ (right column) wavelength for different inclination angles. Images at wavelengths $\lesssim 8\mic$ show brightness behaviour similar to this example. A faint brightness ring in images at $\sim 2R_{in}$ distance from the centre is a feature caused by $0.1\mic$ grains in the disc surface. The flux through each image pixel is scaled with the bolometric flux $F_{bol}$. Contributions from $0.1\mic$ and $2\mic$ dust grains to the model images is shown in Fig.~\ref{Fig_images_grains}.}
  \label{Fig_images}
\end{figure*}

\begin{figure*}
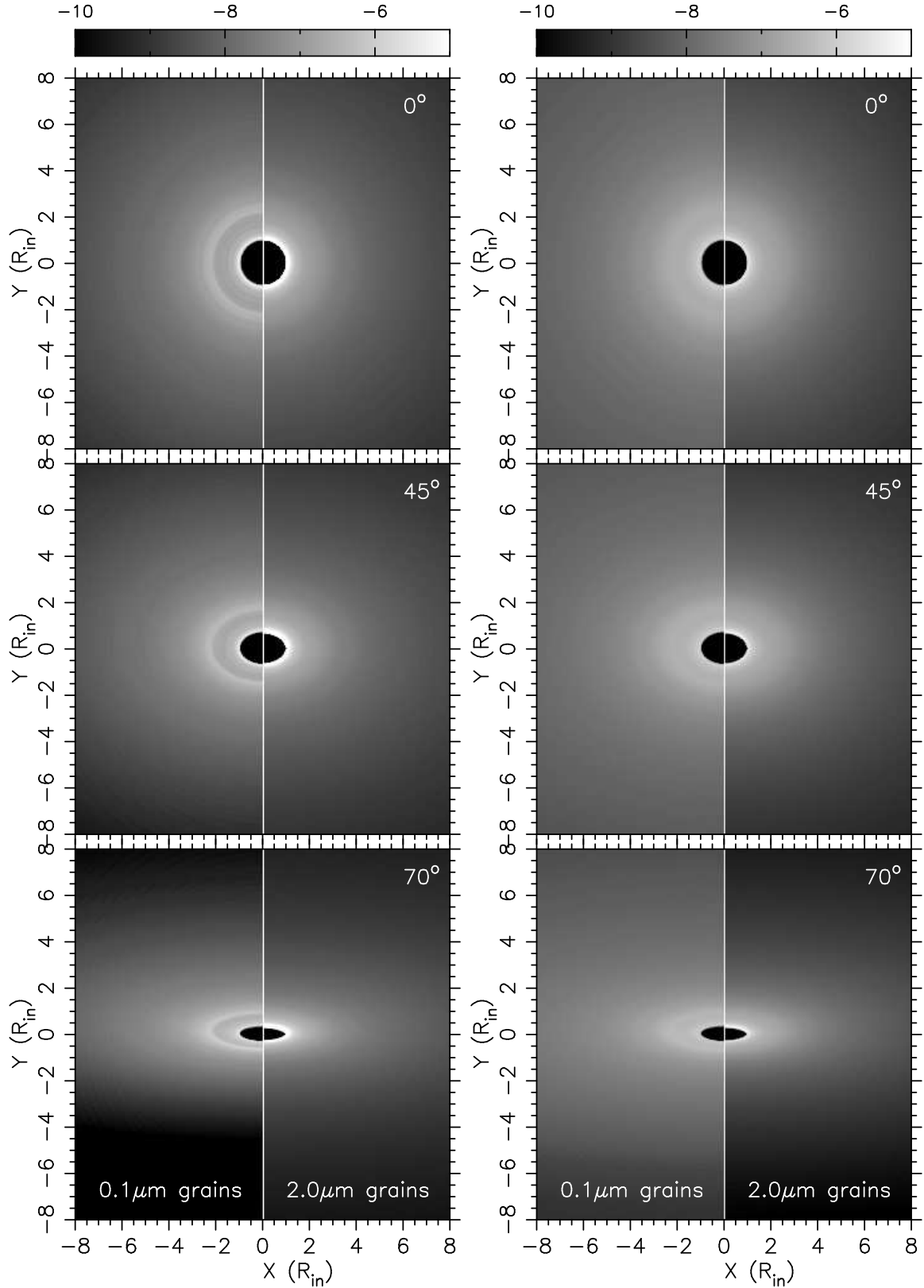

  \includegraphics[width=3in]{Figure_11a.ps}
  \includegraphics[width=3in]{Figure_11b.ps}\\
  \caption{Contribution from $0.1\mic$ and $2\mic$ dust grains to the model images at $2.2\mic$ (left column) and $10\mic$ (right column) wavelength for different inclination angles. The left side of images shows contribution from $0.1\mic$ dust grains and the right side from $2\mic$ dust grains. The flux through each image pixel is scaled with the bolometric flux $F_{bol}$. For the complete images merging both contributions see Fig.~\ref{Fig_images}.}
  \label{Fig_images_grains}
\end{figure*}

\begin{figure*}
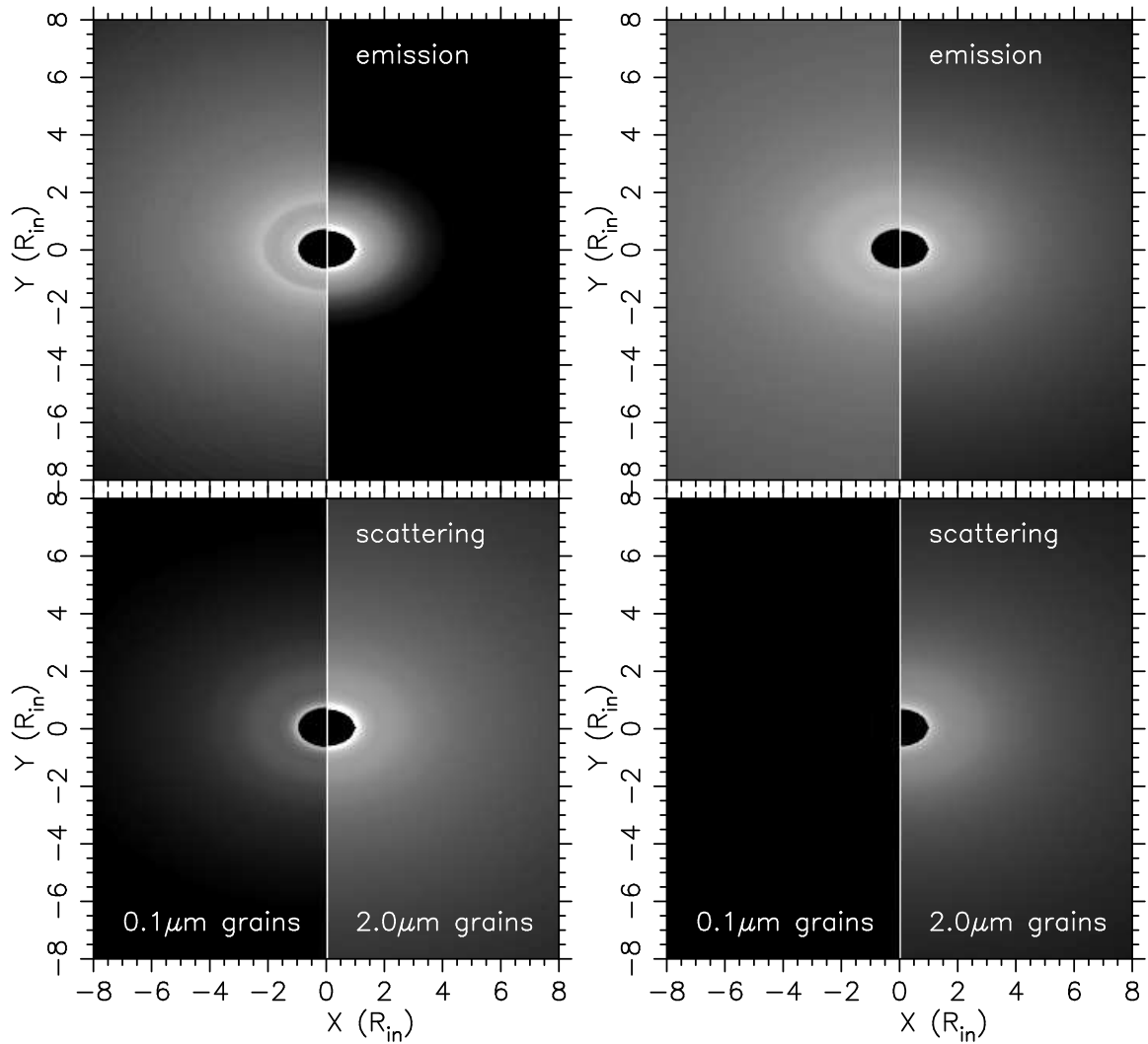

  \includegraphics[width=3in]{Figure_12a.ps}
  \includegraphics[width=3in]{Figure_12b.ps}\\
  \caption{Model image at $2.2\mic$ (left column) and $10\mic$ (right column) wavelength and $45^\circ$ inclination separated into its dust scattering and emission component. The left side of images shows contribution from $0.1\mic$ dust grains and the right side shows contribution from $2\mic$ dust grains. The total image is shown in Fig.~\ref{Fig_images}.}
  \label{Fig_image_em_sca}
\end{figure*}

\subsection{Spectra and images}

The SED of protoplanetary discs traces the dust emission and scattering at different temperatures and optical depths. The inner disc region experiences the highest temperatures and it should be detectable at near IR and mid IR wavelengths. Large optical depths prevent photons of these wavelengths from the deep disc interior to be detected; hence, the optically thin disc surface dominates the flux. Fig.~\ref{Fig_spectra} shows SEDs of the model viewed face on (inclination angle $\theta=0^\circ$) and inclined ($\theta=45^\circ$ and $\theta=70^\circ$). The SEDs do not differ at short wavelengths where the stellar flux dominates. They also do not differ at the far IR wavelengths where the disc becomes optically thin to its own emission.

The relative contribution of stellar flux, dust emission and scattering to the total SED is shown in Fig.~\ref{Fig_fractions} for the face on disc model. Inclination decreases scattering contribution and increases the contribution of dust emission, but the overall wavelength variation is similar to this face on example. The competition between dust emission and scattering for the dominance in the infrared part of spectrum is closely matched with their interplay in the dust opacity shown in Fig.~\ref{Fig_cross_sec}. The dominance of big grains in the disc surface leads to strong contribution of dust scattering in the near and mid IR. The details depend on the chosen dust properties, but the overall trend indicates that scattering is a significant contributor to the SED.

The near and mid IR part of the SED decrease approximately as $\cos\theta$ for small or intermediate angles. The surface is geometrically thin and $\cos\theta$ comes from the geometrical projection of the emitting surface. For larger inclination angles the SED decreases slower than $\cos\theta$, since the line of view is grazing the disc surface and its geometrical thickness is not negligible any more. A somewhat different SED behaviour is visible at $2-3\mic$ where the flux is almost independent of inclination for small angles and then decreases only slightly with larger angles. These are wavelengths where the inner disc edge emits at $T_{sub}$ temperature. The curvature and vertical height of the inner disc edge enable its emission to be almost insensitive to $\theta$, until the inclination becomes large \citep{Isella06, Tannirkulam07, Thi}.

Model images shown in Fig.~\ref{Fig_images} demonstrate this. The inner disc edge is the brightest feature in $2.2\mic$ images and inclination exposes its curvature. At $10\mic$ the disc edge is not the main feature and the rest of the disc surface dictates the SED. Disentangled contribution from both grains sizes to the disc images is shown in Fig.~\ref{Fig_images_grains}. The brightness of the inner disc edge is dominated by big grains, but smaller grains also contribute despite being positioned optically deeper within the disc. A faint ring visible at $\sim 2.2$ distance from the image centre is the location where smaller grains show up in the disc surface.

The near IR images and SED are obviously highly sensitive to the curvature of the inner disc edge. This is why its height and shape has been of great interest to astronomers \citep{Millan-Gabet07,ReviewARAA}. But the rest of the disc surface also contributes to the image flux. In Fig.~\ref{Fig_image_em_sca} we show how dust scattering and emission differ in their contribution to the disc surface brightness at $2.2\mic$ and $10\mic$ wavelength. The thermal emission is modulated by the dust emissivity and reddened by dust layers along the line of sight. Fig.~\ref{Fig_image_2.2mic_temperature} shows the temperature distribution across the $2.2\mic$ image at $\tau=1$ optical depth, which approximately shows how deep we see into the disc. From Fig.~\ref{Fig_image_em_sca} we see that temperatures of $2\mic$ grains contributing the most to the thermal emission at near IR wavelengths are limited to small radii close to the inner disc edge, while $0.1\mic$ grains contribute over larger area, but with smaller surface brightness due to their smaller emissivity. In the mid IR both grains have thermal emission spread over a larger area. Since $0.1\mic$ grains appear in the disc surface layer at $\varrho\sim 2.2$, we can see a brightness increase in images at this distance from the centre. However, this increase is small because the emissivity of $0.1\mic$ grains is small compared to $2\mic$ grains. This is why big grains dominate the SED and images in the near IR even though smaller grains have high temperatures over a larger portion of the disc surface.

While the emission is confined to small radii where the dust is the hottest, scattering follows surface dust density and extends further away. \cite{Pinte} showed how scattering is an important component in the interpretation of the near IR imaging data. We need to emphasize here that our images are calculated with isotropic scattering function, which is a rough approximation for big grains. A more realistic scattering function would yield larger image asymmetries with inclination.

\section{Discussion and Conclusions}
\label{sec_discuss_conclude}

In this paper, we have explored the temperature structure of the hottest dusty regions of protoplanetary discs. In this the most inner part of the disc, the interplay of dust density, optical depth and multigrain dust composition creates an intricate disc surface structure shaped by nonuniform sublimation of individual dust species. Our study was limited to two distinctively different dust grain radii of $0.1\mic$ (small) and $2\mic$ (big). These grains exhibit very different sublimation behaviour when used individually in single grain models, with big grains surviving much closer to the star than small grains. However, we demonstrated a significant decrease of sublimation distance of small grains when they are mixed with big grains under conditions of big grains dominating the opacity. Small grains survive behind big grains' visual optical depth of $\tau_V\sim 4-6$, which translates into only a few percentages larger sublimation distance than big grains.
We argue that the disc accretion constantly resupplies the inner disc with big grains, which enables them to dominate the opacity and sustain conditions described in our paper. There are also observational confirmations of this inner disc property; spectra indicate large fractions of big grains in the surface of inner discs \citep[e.g.][]{vanBoekel04,vanBoekel05,Sargent09,Watson09,Juhasz10}, while the observed sizes of inner discs are consistent with big grains dictating the inner disc radius $R_{in}$ \citep{Millan-Gabet07}.

Our study required very high self-adaptive computational grid resolution (almost $10^{-6}R_{in}$) in order to correctly trace dust sublimation and temperature gradients. Computations have to resolve two important temperature gradients. One is the temperature inversion experienced by big grains, where the temperature reaches its maximum at $\tau_V\sim 1.5$ instead of the very surface exposed to the stellar radiation. The other comes from the tendency of small grains to quickly thermalize with big grains as the optical depth is increasing, which results in a steep temperature decline for small grains. Prior similar studies were not resolving these effects in multigrain dust models \citep{Tannirkulam07, Kama}, which signifies the importance of high resolution radiative transfer calculations.

Implications of the temperature inversion effect, where the dust temperature at $R_{in}$ is lower than sublimation, have received modest attention in the literature so far \citep{Vinkovic06,Kama,ReviewARAA}. The main consequence of this effect is that big grains can survive closer to the star than optically thick part of the disc. For theorists this implies that currently used dusty disc models are incomplete, while developing a more realistic model will require scrutinizing the dust dynamics coupled with radiative transfer. For observers this implies that optically thin dust might be detectable within the currently used optically thick inner disc radius. Recent detection of disc brightness within $R_{in}$ could be the first indication that such a dust structure exists, but it has to be disentangled from possible gas emission \citep{Tannirkulam08,Benisty10}.

We also resorted to a simplified boundary condition where the dust density drops to zero at $R_{in}$, simply because we currently do not have any theoretical handle on the density structure of optically thin dust within $R_{in}$. Its structure can be highly influenced by dust dynamics due to other forces than gravity and gas drag. Radiation pressure force would act upon all dust grains in this optically thin disc zone. Magnetic fields at these distances could be strong enough to interact with charged dust particles. Charging is expected because of high gas temperatures and exposure to UV and X-ray radiation \citep{Turner,Pedersen}. This leads to coupling of grain settling to the distribution of the magnetorotational turbulence \citep{Turner}. High gas densities also provide conditions for photophoresis on big grains, which moves particles outward \citep{Moudens}. Moreover, these dynamical processes could produce optical depth variability within $R_{in}$, which would immediately influence the optically thick region where the dust at the sublimation zone can overheat and sublimate. Since this sublimation zone is located within a disc, we do not know how optical depth changes caused by dust sublimation would influence the overall disc structure. All these issues make further work on the temperature structure of inner disc region a very challenging task \citep{ReviewARAA} that will require computation of individual trajectories of dust grains coupled with radiative transfer. Such models are still under development for studying the outer, dynamically and thermally simpler disc regions \citep[e.g.][]{Charnoz}.

We have also analyzed how differences in the distribution of small and big grains influence the disc spectrum and images. Big grains dominate the near IR spectrum and image flux, since small grains cannot survive in the disc surface up to distances $\gtrsim 2.2 R_{in}$. Nonetheless, small grains contribute significantly to the image surface brightness, especially at scales several times larger than $R_{in}$. Hence, invoking grain growth as an interpretation for a larger fraction of big grains in the inner part of protoplanetary discs has to take into account the dominance of big grains in the near IR opacity of dust mixture and the spatial separation of dust grains by dust sublimation. Notice that in our model we used the same ratio of big to small grains through the entire disc, but, nevertheless, their fractional contribution to the SED and images is highly wavelength dependent.

Dust separation is also caused by radiation pressure force, which we will explore in the next paper. Small grains are more easily eroded from the surface of optically thick discs than bigger grains due to the radiation pressure from stellar radiation \citep{TakeuchiLin}. Moreover, \cite{Vinkovic09} showed how big grains experience additional radiation pressure force due to the near IR radiation from the hot disc dust, which pulls bigger grains out of the disc until the stellar radiation pressure force overcomes the gas drag force and blows grains outward. This process is very fast (within orbital time) unless the local gas density is high enough to dominate over radiation pressure force. Another important effect is corrections to the vertical size of the inner disc. The inner disc edge is hotter than the standard disc structure, which results in vertical expansion of the disc \citep{DDN, Thi}. As long as the disc does not expand so much to cause self-shadowing, we expect its temperature structure to be qualitatively similar to our model, except for geometrical scaling of the vertical disc size. Another important thermal effect comes from viscous heating, which increases the temperature in the midplane. However, here we again expect the thermal structure above the midplane to be dominated by effects described in our model, since this region is dominated by stellar heating \citep{DAlessio}. Overall, the exact dynamical stability of dusty disc structure taking into account all these effects, as well as optically thin dust within $R_{in}$, is still unknown and it requires additional research.

\section*{Acknowledgments}

The project described in this paper was performed over many years using various computational facilities. It was supported by National Computational Science Alliance (NCSA) under AST 04-0006 and utilized the NCSA's Xeon Linux Cluster. The author also thanks the Institute for Advanced Study for time on their Linux cluster and the University Computing Centre SRCE in Zagreb for time on their cluster Isabella. Finally, the completion of the work was performed on computer cluster Hybrid at the Physics Department, University of Split, financed by the National Foundation for Science, Higher Education and Technological Development of the Republic of Croatia.

\end{document}